\documentclass[preprint]{aastex63}
\usepackage{graphicx}
\usepackage{amssymb,amsmath}
\hypersetup{linkcolor=red,citecolor=blue,filecolor=cyan,urlcolor=magenta}
\received{\today}
\shorttitle{Does a filament barb always correspond to a filament foot?}
\shortauthors{Ouyang et al.}

\begin{document}

\title{Does a solar filament barb always correspond to a prominence foot?}

\correspondingauthor{P. F. Chen}
\email{chenpf@nju.edu.cn}

\author{Y. Ouyang}
\affiliation{School of Physics and Electronic Engineering, Linyi University, Linyi 276000, China}
\affiliation{Key Lab of Modern Astron. \& Astrophys. (Ministry of Education), Nanjing University, China}

\author[0000-0002-7289-642X]{P. F. Chen}
\affiliation{School of Astronomy \& Space Science, Nanjing University, Nanjing 210023, China}
\affiliation{Key Lab of Modern Astron. \& Astrophys. (Ministry of Education), Nanjing University, China}

\author{S. Q. Fan}
\affiliation{School of Physics and Electronic Engineering, Linyi University, Linyi 276000, China}

\author{B. Li}
\affiliation{School of Physics and Electronic Engineering, Linyi University, Linyi 276000, China}

\author{A. A. Xu}
\affil{State Key Laboratory of Lunar and Planetary Sciences, Macau University of Science and Technology, Macau, China}

\begin{abstract}
Solar filaments are dark structures on the solar disk, with an elongated spine and several barbs extending out from the spine. When appearing above the solar limb, a filament is called a prominence, with several feet extending down to the solar surface. It was generally thought that filament barbs are simply the prominence feet veering away from the spine and down to the solar surface. However, it was recently noticed that there might be another dynamic type of barbs, which were proposed to be due to filament thread longitudinal oscillation. If this is the case, the dynamic barbs would not extend down to the solar surface. With the quadrature observations of a filament barb on 2011 June 5 from the {\it Solar Dynamics Observatory} and the {\it STEREO} satellites, we confirm that the filament barb is due to filament thread longitudinal oscillations. Viewed from the side, the filament barb looks like an appendix along the spine of the prominence, and does not extend down to the solar surface as a foot.
\end{abstract}


\section{Introduction}

Solar filaments are a remarkable phenomenon in the solar atmosphere. They are elongated structures, with cold dense plasma suspended in the hot tenuous corona. Hence, when appearing near the solar limb, they lie high up in the corona and are therefore called prominences. In the past decades, they have attracted wide attention since, once erupting, they are strongly associated with solar flares and coronal mass ejections \citep{che11, par14, sch13}. On the solar disk, a solar filament is generally seen as an elongated structure called spine, with or without several barbs veering away from the spine \citep{kie53}. Barbs are one basic structural component of solar filaments \citep{mar94, mac10, mar18}. They extend laterally from the filament spine, like ramp roads of a highway \citep{mar94, pev03, mac10, che14, hao15, ouy17}. When a filament appears above the solar limb, where it is called prominence, it exhibits one or more feet at the sites of the original filament barbs, extending down from the spine to the solar surface, like pillars seemingly supporting the prominence \citep{oka07, mac10, arr12, liu12, par14,zha16}. As a result, it was generally thought that there is one-to-one correspondence between filament barbs and prominence feet.

As a basic structural component of solar filaments, barbs are not randomly distributed along the filament channel. First, they are usually rooted near the parasitic polarity in the photospheric magnetograms. Second, they present systematic pattern known as filament chirality \citep{mar94}, i.e., filament barbs can be classified to be either left-bearing or right-bearing types. It was further proposed that there is a one-to-one relationship between the filament chirality and the handedness of the magnetic topology of the filament channel or between the filament chirality and the handedness of the magnetic arcade overlying the filament channel \citep{mar94}. As illustrated by \citet{mar98}, the left-bearing barbs would imply positive helicity of the filament channel, and the right-bearing barbs correspond to negative helicity of the filament channel. Such a rule has been widely used to infer the helicity sign of filament channels, either in individual events \citep{suj05, chan10, bal11, jia14} or in large samples \citep{mar98, pev03, yea07, hao15, haz18}. However, this one-to-one correspondence was challenged by \citet{guo10} and \citet{che14}, who pointed out that such a one-to-one correspondence is valid only for the magnetic flux rope configuration, and fails in the sheared-arcade magnetic configuration. Furthermore, \citet{che14} proposed that, instead of revealing the helicity sign of a filament channel, filament barbs, in combination with the helicity sign, can be used to infer the magnetic configuration, i.e., flux rope or sheared arcade, where the helicity sign can be determined by the conjugate brightenings produced by the filament drainage or coronal cells \citep{wang09, shee12}. It is noted in passing that in case there exist anomalous barbs in one filaments, e.g., due to strong parasitic polarity \citep{fili17}, it was proposed that the bearing sense of filament threads is more reliable in identifying the bearing sense of filaments \citep{mar08}.

Despite their importance, the nature of filament barbs are not well understood, in particular their magnetic configuration remains elusive. Several models have been put forward. Based on the fact that many barbs are related to parasitic polarity, \citet{mar94} proposed a wire model, where barbs correspond to the magnetic field lines linked to the parasitic polarity \citep[see][]{mar94, mar98, mar09, mar18}. There are several issues that seem hard to be explained by this model, for example, the support of the filament feet against gravity and the opposite signs of helicity between the filament and the overlying magnetic arcade. On the other hand, barbs are considered to be supported by an ensemble of magnetic dips due to the intrusion of parasitic polarity \citep{aul98, cha05, mac09}. With linear or nonlinear force-free magnetic field modeling, \citet{aul98} and \citet{mac09} found that the existence of parasitic polarity can reproduce many observational features of filament barbs, e.g., barbs generally terminate near the boundary of both minority and dominant polarity elements \citep{wan01, lin05}. In their models, the barbs seen from the top indeed correspond to the filament feet seen from the side.

Although it has been widely taken for granted that filament barbs correspond to filament feet, it is arguably doubtful that the correspondence is not universal. First, similar to \citet{aul02}, \citet{van04} found in one filament that the barb is not related to any parasitic polarity at all. Instead, it is due to the extension of the magnetic dips away from the main path of the filament caused by weak dominant polarity at the photosphere. As claimed further by \citet{che14}, the real photospheric magnetic field is always nonuniform, which would naturally lead to different lengths of magnetic dips along one filament channel and hence filament threads \citep{zho17}. The outstanding ones would be recognized as filament barbs. In these cases, a barb is simply a horizontal extension of a segment of filament spine, and bald-patch magnetic configuration does not exist below the barb, hence there is no corresponding foot appended to the filament spine. Second, recently \citet{awa19} proposed dynamic barbs, which are due to longitudinal oscillations of filament threads. In this scenario, it is expected that the filament barbs do not correspond to any feet at all.

In order to confirm that the dynamic barbs do not correspond to any filament feet, the best way is to observe a dynamic barb from orthogonal viewing directions, i.e., one from the top in order to identify the barb and the other one from the side in order to check the absence of any filament foot. Fortunately, the {\em Solar Terrestrial Relations Observatory} ({\it STEREO}) twin satellites and {\em Solar Dynamics Observatory} ({\it SDO}) provided us excellent chances to conduct such quadrature observations. In this paper, we found one filament, which was simultaneously observed by {\it SDO} and {\it STEREO} in a quadrature manner. The observations are briefly described in \S\ref{sec2}, and the results are presented in \S\ref{sec3}, which are followed by discussions in \S\ref{sec4}.

\section{Observations and data analysis}\label{sec2}

On 2011 June 5, a filament was observed to be near the central meridian in the southern hemisphere in the field of view of the Extreme Ultraviolet Imager (EUVI), which is on board the {\it STEREO}-Behind, or {\it STEREO-B} hereafter. At that time, {\it STEREO-B} spacecraft was separated from the Earth by an angle of $\sim$93$^\circ$, facilitating the quadrature observations with the telescopes orbiting the Earth. The EUVI \citep{wue04} is one instrument of the Solar Earth Connection Coronal and Heliospheric Investigation (SECCHI) suite on board the {\it STEREO} twin satellites. It observes the solar disk with several EUV wavelengths, including 304 \AA, 171 \AA, 195 \AA, and 284 \AA. In this paper, we use the 195 \AA\ images to investigate the dynamics of the filament and its barbs. The cadence of the observation is 5 minutes, and the pixel size of each image is $1\farcs59$.

Such a filament in the {\it STEREO-B} field of view was observed as a prominence near the solar east limb from the vantage point of {\it SDO}. One of the three onboard telescopes, Atmospheric Imaging Assembly \citep[AIA,][]{lem12}, observes the Sun in seven EUV and three UV channels with a pixel size of $0\farcs6$ and a time cadence of 12 s. In this study we use the 193 \AA\ bandpass in order to match the observation of {\it STEREO-B}/EUVI 195 \AA\ channel.

In order to study the dynamics of the filament/prominence, all the {\it STEREO-B}/EUVI images are derotated to the universal time 19:50:55 UT. All the EUV data are processed with the corresponding SSWIDL software.

\section{Results}\label{sec3}

During its passage across the central meridian in the {\it STEREO-B}/EUVI field of view, the filament is activated episodically, as indicated by the animation associated with Figure \ref{fig1}. The upper panels of Figure \ref{fig1} display 4 snapshots of the filament evolution observed by {\it STEREO-B}/EUVI. It is seen that at the 19:50 UT (panel a), only the elongated main body of the filament is clearly seen. No pronounced barbs can be recognized. At 20:05 UT (panel b), a significant barb with a length of 32$\arcsec$ extends from the left side of the southern part of the filament around the latitude of south $21^\circ$, as indicated by the white arrow. This barb is left-bearing, which is the dominant chirality of filaments in the southern hemisphere. The barb makes an angle of $\sim$25$^\circ$ relative to the spine of the filament. At 21:30 UT (panel c), while the barb on the left side of the filament is receding, another barb on the right-hand side of the filament stretches out from the spine, which is aligned with the left-hand barb in panel (b), as marked by the white arrow in panel (c). This barb recedes as well, becomes invisible at 22:20 UT (panel d), continuing its appearance on the left side and right side alternatively.

Such an evolution is simultaneously observed by the {\it SDO}/AIA. Its 193 \AA\ intensity maps at 4 snapshots closest to the corresponding times of the {\it STEREO-B}/EUVI observations are displayed in the lower panels of Figure \ref{fig1}. We can see that whenever a barb is detected by the {\it STEREO-B}/EUVI (panels b--c), a thread stretches out from the main body of the prominence, as indicated by the arrows in Figure \ref{fig1}(f--g). Interestingly, the stretching thread is higher than the top of the prominence main body like two tentacles stretching out toward the north and south directions. The length of the tentacle is about 29$\arcsec$ in Figure \ref{fig1}(f) and 42$\arcsec$ in Figure \ref{fig1}(g). It is noted that due to the tilted rotation axis of the Sun and the different vantage angles of the two satellites, the heliocentric coordinates of the filament/prominence are different between the {\it STEREO-B}/EUVI and {\it SDO}/AIA maps. However, the latitudes of the filament/prominence in the two satellites are the same, i.e., between 16.3 and 31.8 degrees in the southern hemisphere, and the barbs in Figure \ref{fig1}(b--c) and the tentacles in Figure \ref{fig1}(f--g) all lie in the same latitude zone, i.e., around 21 degrees in the southern hemisphere as marked by the dashed lines.

From both the top and side views, it is revealed that the filament barbs in our event stretch out and retract alternatively, reminding us of the filament thread longitudinal oscillations. In order to confirm it, we select two slices along the filament barb in Figure \ref{fig1}(c) and the prominence tentacle in Figure \ref{fig1}(g), respectively. The width of the slices in both Figure \ref{fig1}(c) and Figure \ref{fig1}(g) is taken to be 2 pixels. The time evolution of the EUV intensity along the two slices are plotted in the two panels of Figure \ref{fig2}, where the top panel corresponds to the time--distance diagram for the {\it STEREO-B} barb, and the bottom panel to that for the {\it SDO}/AIA tentacle. The selected time range is between 18:20 UT on June 5 and 01:52 UT on June 6, and both the barb and the tentacle are absorptive structures. It is seen that during this episode the filament activation starts before 20:00 UT. While the filament barb is visible in the top panel before 20:00 UT, the prominence tentacle is visible only after 20:00 UT. Both the barb structure and the tentacle structure start to oscillate, with the amplitude decaying with time. The oscillations of both the filament barb and the prominence tentacle last about three and a half periods.

In order to compare their oscillations quantitatively, we fit their oscillating patterns with decayed sine functions as done by \citet{zha12},
\begin{equation}\label{eq1}
y =A\cos (2\pi t/P + {\phi}_0){e}^{-t/\tau} + y_0,
\end{equation}
\noindent
where $A$ represents the initial amplitude, $P$ is the period, $t$ is the time lapse since 21:00 UT, $\phi_{0}$ is the initial phase angle, $\tau$ is the decay time, and $y_{0}$ is the equilibrium position. Since the first half period of the oscillation is apparently different from other periods in both panels, we fit the absorption structures only during the last three periods. It is found that the intrinsic parameters of the oscillation, e.g., the period and the decay time, are the same for both panels, with $P=1.2$ hr and $\tau=2$ hr. And even the initial phase angle is also the same for both panels, i.e., $\phi_{0}=0.31$. Only the other non-intrinsic parameters are different for the two panels, i.e., $A=26\arcsec$ and $y_0=34\arcsec$ for {\it STEREO-B} and $A=60\arcsec$ and $y_0=55\arcsec$ for {\it SDO}/AIA. The fitted profiles are overplotted in Figure \ref{fig2} as white dotted lines. The three peaks of both panels match very well as indicated by the dashed-dotted lines, i.e.,  the {\it STEREO-B} barb and the {\it SDO}/AIA tentacle oscillate synchronously all the time. It is also seen that the oscillations between 20:00 UT and 22:00 UT do not match the fitted profiles, indicating the oscillation period is longer initially. It is interesting to see that even during this interval the {\it STEREO-B} barb and the {\it SDO}/AIA tentacle are synchronous.

\section{Discussions}\label{sec4}

With high-resolution observations, it has been shown that both barbs and the spine (including the two ends) of a filament is composed of many threads \citep{lin08, mac10, yan15}, which are skewed from the filament spine by 3--30 degrees \citep{ath83, hir85, ler89, han17}. It is generally believed that these threads follow the local magnetic field lines \citep{eng79, mar08, yan15}, where the cold plasma is supported against gravity by the local magnetic dips in most cases. Once perturbed by external disturbances, the threads inside a filament would be pushed aside from the equilibrium positions, and then start to oscillate under the restoring force. Filament oscillations can be classified into transverse oscillations and longitudinal oscillations \citep{arr18}, depending on whether the perturbation is perpendicular or parallel to the local magnetic field \citep{zho18, zha19, adr20}. The two modes may coexist in one event \citep{wan16, zha17, maz20}. In terms of longitudinal oscillations, there might be collective oscillations or local oscillations: If the disturbances are in large scale, e.g., an incident EUV wave \citep{she14}, all the threads inside the filament might oscillate collectively, forming filament longitudinal oscillations as a whole; If the disturbances are due to localized heating or plasma ejection\citep{jin03, vrs07, che08, zha12, lun14}, only a bundle of threads might oscillate along the local magnetic field. In this case, the oscillating thread bundle would form a dynamic barb, which is skewed from the filament spine by 3--30 degrees.

\begin{figure*}
\epsscale{1.0}
\plotone{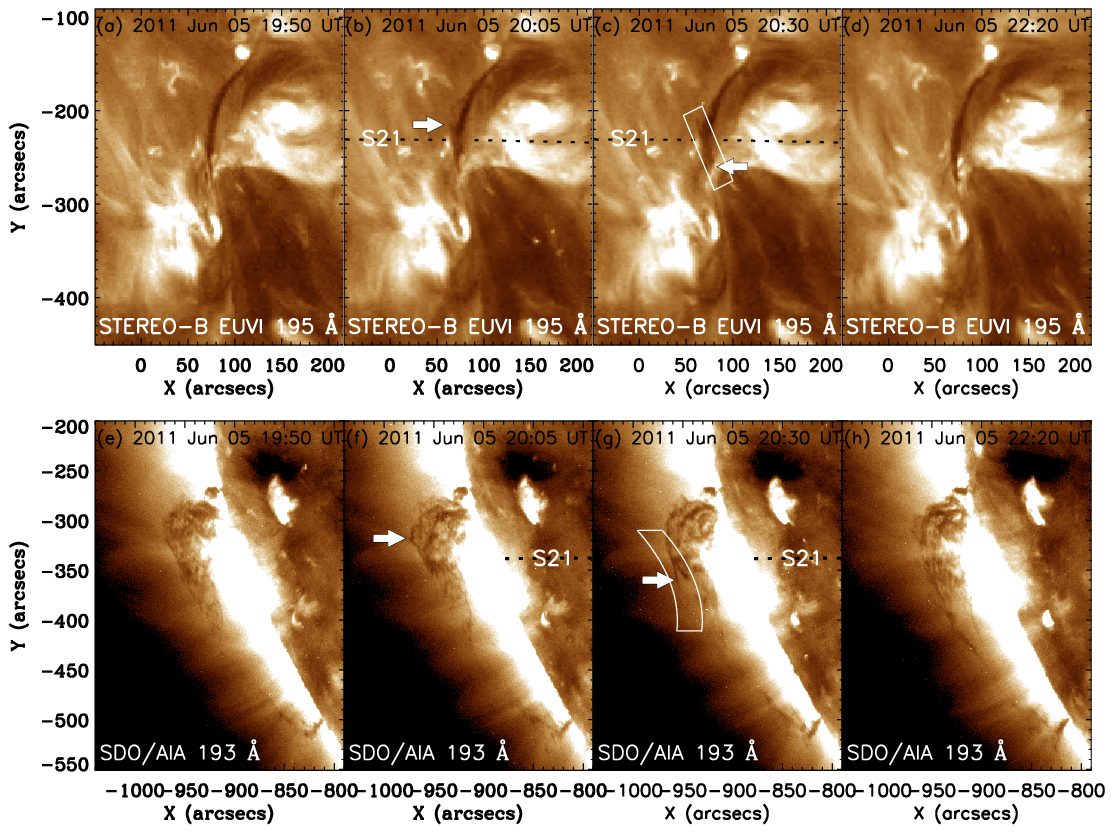}
\caption{Four snapshots of the filament evolution observed by the {\it STEREO-B}/EUVI at 195 \AA\ waveband (top panels) and by the {\it SDO}/AIA at 193 \AA\ waveband (bottom panels) on 2011 June 5, when the {\it STEREO-B} spacecraft was separated from the Earth by an angle of $\sim$93$^\circ$. In panels (b, c, f, and g), the arrows indicate the dynamic barbs, the black dashed lines indicate the latitude of south 21$^\circ$. Two slices marked by the white solid boxes are selected for plotting the time-distance diagram in Fig. \ref{fig2}.}
	\label{fig1}
\end{figure*}

\citet{awa19} reported this kind of dynamic barbs, and proposed that they are due to filament longitudinal oscillations. In order to confirm it, we investigated the dynamics of a filament barb on 2011 June 5 with quadrature observations, one from the top and one from the side. For the following reasons, the filament barb indeed corresponds to the longitudinal oscillations of a bundle of filament threads. First, as indicated by Figure \ref{fig1}(c) and the associated animation, the filament barb moves along the direction of the thread, forming a barb skewed from the filament spine by an angle of $\sim$25$^\circ$. Such a value is in the typical range of the angle between the filament spine and the supporting magnetic field \citep{ler84, lun14}. Second, when the filament threads stretch out to form the dynamic barb in the top view of {\it STEREO-B}/EUVI, the side view of {\it SDO}/AIA revealed that a structure similar to a tentacle emanates from the spine. Both the {\it STEREO} barb and the {\it SDO} tentacle are at the same latitude, and oscillate synchronously. The oscillation period is found to be 1.2 hr in the final 3 cycles, which is a typical value for filament longitudinal oscillations \citep{jin03, vrs07, che08, zha12, lun14}. Since filament threads are situated at the lowest portion of magnetic dips, it is expected to see that they ascend to a higher altitude as they deviate from the equilibrium positions. This is exactly what we observed in Figures \ref{fig1}(f--g). Third, as the filament barb in the {\it STEREO} view stretches out, no foot appears below the prominence spine. In order to illustrate it, we plot one snapshot of the {\it SDO}/AIA base-difference image in Figure \ref{fig3}, which correspond to the dynamic barb which just stretches out at 20:30 UT. The base time is chosen to be 20:15 UT. It is revealed in Figure \ref{fig3} that only the dynamic tentacle is present, and no any foot extends down from the prominence spine to the solar surface at the location of the prominence tentacle. The absence of a prominence foot in 193 \AA\ might also be due to the variation of plasma temperature. However, after checking a colder waveband (171 \AA) and a hotter waveband (211 \AA), we found that no prominence foot is present below the prominence tentacle in all {\it SDO}/AIA wavebands, indicating that the absence of a prominence foot is real.

The dynamic barbs are similar to the traditional barbs in many senses. For example, considering a typical longitudinal oscillation with a period of 1 hr and the maximum velocity of 20 km s$^{-1}$, the length of the dynamic barb would be 11 Mm, which is near the lower limit of the longest traditional barbs \citep{li13}. Besides, similar to the traditional barbs whose bearing orientation tells the chirality of the filament, the dynamic barbs are formed due to filament thread longitudinal oscillations, therefore their orientation follow that of the threads. As pointed out by \citet{mar08}, the thread orientation is even better than traditional barbs in characterizing the chirality of filaments. In this paper, the dynamic barb indicates the filament has a left-bearing chirality, which is the dominant one in the southern hemisphere \citep{mar98, pev03, ouy17}. More importantly, the traditional barbs correspond to the lateral feet of the filament extending down to the solar surface. Such a 3-dimensional structure strongly suffers from the projection effects, i.e., when a filament is far from the solar disk center, its left-bearing barbs might be falsely identified as being right-bearing, and vice versa. However, dynamics barbs extend mainly along the horizontal magnetic field along the filament thread, the projection effects are significantly weaker.

On the other hand, the dynamic barbs are different from the traditional barbs in several aspects. First, the traditional barbs remain roughly unchanged for hours or tens of hours, although the threads in these barbs were disclosed to oscillate slightly as well \citep{li13}. However, the dynamic barbs protrude and retract periodically with a period of $\sim$1 hr, and decay in $\sim$3--4 periods. Second, since dynamic barbs are due to filament thread longitudinal oscillations, they are not manifested as lateral feet of the prominence, and they have nothing to do with the parasitic polarity in the photospheric magnetogram.

Having talked about the filament dynamic barbs that do not correspond to filament feet, it is interesting to note in passing that even the traditional quiescent barbs are not necessary to be the filament feet, and are not supported against gravity by bald-patch magnetic configurations. As mentioned by \citet{che14}, due to the nonuniformity of the photospheric magnetic field, the magnetic dip series along the filament spine presumably have different lengths, which would result in filament threads with different lengths. When the photospheric magnetic field in the filament channel is extremely nonuniform, a bundle of threads is significantly longer than others, which would be manifested as a filament barb as modeled by \citet{van04}. Although it was not explicitly mentioned in their paper, it would be expected that the filament barbs produced this way are not due to bald-patch magnetic field, and there are no filament feet in the side view.

\begin{figure*}
\epsscale{1.0}
\plotone{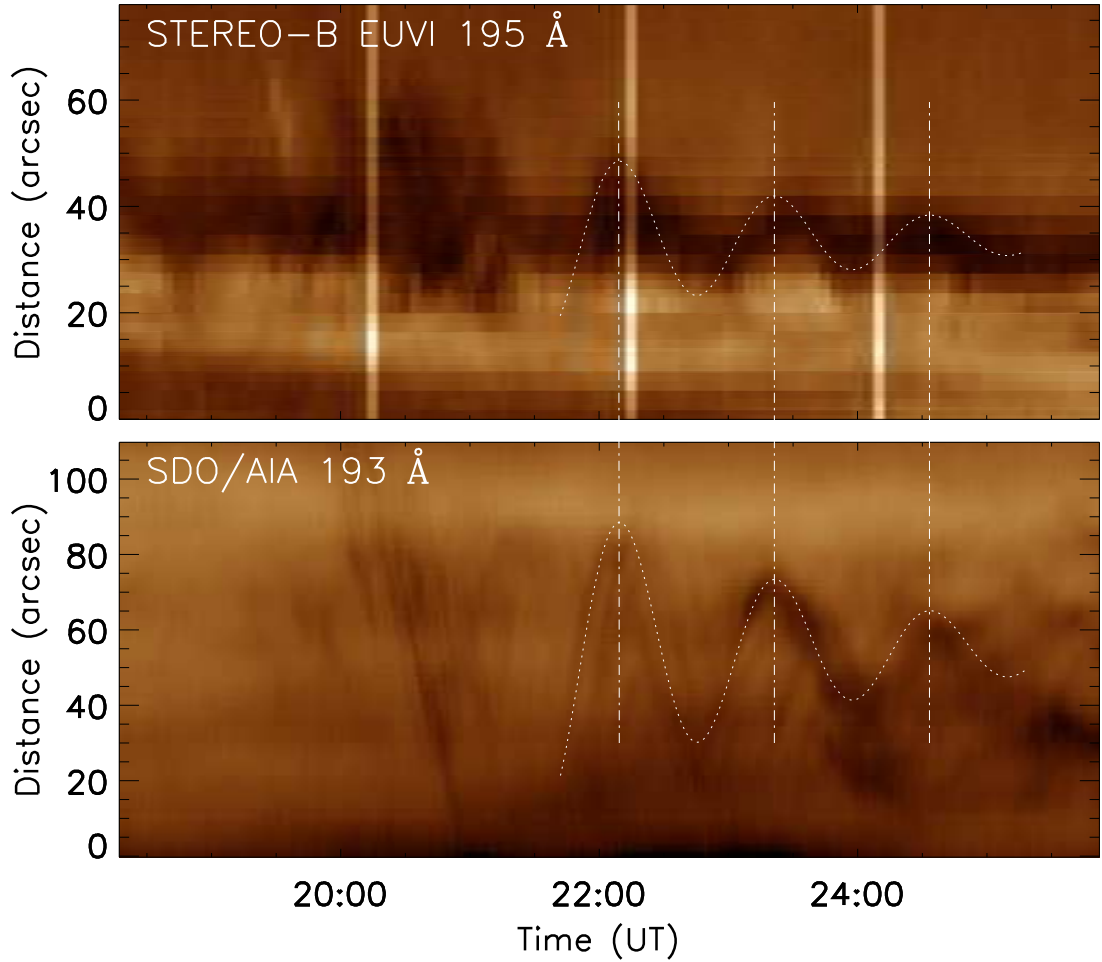}
\caption{Time-distance diagrams of the {\it STEREO-B}/EUVI 195 \AA\ intensity (top panel) and the {\it SDO}/AIA 193 \AA\ intensity (bottom panel). The corresponding slices are marked in Figs. \ref{fig1}(c) and \ref{fig1}(g), respectively. The filament/prominence is visible as the dark strands, and the oscillating patterns are fitted with decayed sine functions, which are overplotted as the white dotted lines. The vertical dotted-dashed lines highlight the synchrony of the filament/prominence oscillations.}
  \label{fig2}
\end{figure*}

\begin{figure*}
\epsscale{1.0}
\plotone{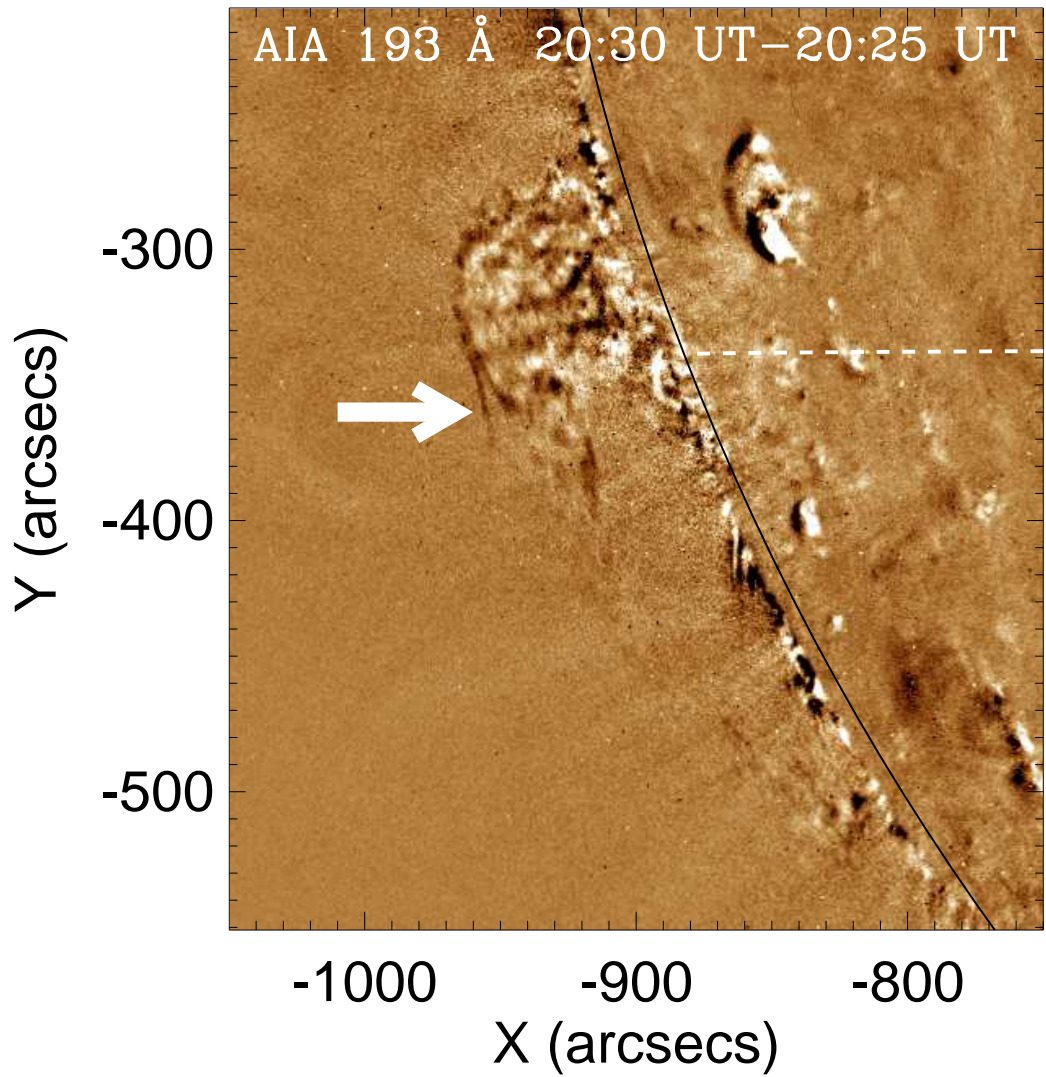}
\caption{ The {\it SDO}/AIA base-difference image at 20:30 UT on 2011 June 5, the base time is 20:25 UT. The arrow indicates the oscillating thread, the black curve marks the solar limb, and the white dashed line corresponds to the latitude of south 21$^\circ$.}
  \label{fig3}
\end{figure*}

There are another two issues worth mentioning. First, the filament thread longitudinal oscillation, which produces the dynamic barbs, has a longer period of $\sim$2 hr in the initial stage, as indicated by Figure \ref{fig2} (from 20:00 UT to 22:00 UT). This is significantly longer than the period of oscillation in the later stage. A similar feature was obtained in 2-dimensional magnetohydrodynamic numerical simulations \citep{zha19}. One possible reason is that the curvature radius of the dipped magnetic field line is not constant along the trajectory, implying that the magnetic field line is the most curved near the lowest position of the magnetic dip. It is noted here that magnetic dips are not always necessary for supporting filaments against gravity, in particular for those filaments composed of flowing threads \citep{wang99, karp01}. However, when filament threads experience longitudinal oscillations as in this paper, magnetic dips should be there so that the field-aligned component of gravity can offer the restoring force. The second issue is about the different amplitudes of the same oscillating object viewed from different directions, i.e., the filament barb and the prominence tentacle. As revealed in Figure \ref{fig2}, the amplitude of the oscillating filament barb in the top panel is much smaller than that of the oscillation prominence tentacle, although they are synchronous. For example, at $\sim$22:10 UT, the amplitude of the oscillating 195 \AA\ filament barb is 14.3\arcsec. However, the amplitude of the oscillating 193 \AA\ prominence tentacle is 33\arcsec. The discrepancy is presumably due to the very difference appearance of the same filament/prominence structure when viewed from different directions, as proposed by \citet{guna18}.

To summarize, in this paper we studied a filament barb on 2011 June 5 with {\it STEREO}/EUVI and {\it SDO}/AIA from quadrature vantage points. It is verified that the dynamic filament barb is due to filament thread longitudinal oscillations. With the analysis we emphasize that filament barbs do not always correspond to filament feet. Some barbs are simply filament threads standing out from the spine roughly in the horizontal plane, rather than extending down to the solar surface.

\acknowledgments
The authors thank the {\it STEREO} and {\it SDO} teams for providing the data. This research was supported by the Chinese foundations NSFC (11533005, 11961131002, and U1731241) and Jiangsu 333 Project. P.F.C. thanks ISSI-Beijing for supporting team meetings on solar filaments.

%\bibliography{reference}

\clearpage

%\clearpage

%\clearpage

\end{document}